# Social influence and the collective dynamics of opinion formation

Mehdi Moussaïd[1*], Juliane E. Kämmer[1], Pantelis P. Analytis[1], and Hansjörg Neth[1]

[1] Center for Adaptive Behavior and Cognition, Max Planck Institute for Human Development, Berlin, Germany

[*] Corresponding author: mehdi.moussaid@gmail.com

## Abstract

Social influence is the process by which individuals adapt their opinion, revise their beliefs, or change their behavior as a result of social interactions with other people. In our strongly interconnected society, social influence plays a prominent role in many self-organized phenomena such as herding in cultural markets, the spread of ideas and innovations, and the amplification of fears during epidemics. Yet, the mechanisms of opinion formation remain poorly understood, and existing physics-based models lack systematic empirical validation. Here, we report two controlled experiments showing how participants answering factual questions revise their initial judgments after being exposed to the opinion and confidence level of others. Based on the observation of 59 experimental subjects exposed to peer-opinion for 15 different items, we draw an influence map that describes the strength of peer influence during interactions. A simple process model derived from our observations demonstrates how opinions in a group of interacting people can converge or split over repeated interactions. In particular, we identify two major attractors of opinion: (*i*) the *expert effect*, induced by the presence of a highly confident individual in the group, and (*ii*) the *majority effect*, caused by the presence of a critical mass of laypeople sharing similar opinions. Additional simulations reveal the existence of a tipping point at which one attractor will dominate over the other, driving collective opinion in a given direction. These findings have implications for understanding the mechanisms of public opinion formation and managing conflicting situations in which self-confident and better informed minorities challenge the views of a large uninformed majority.

## Introduction

In many social and biological systems, individuals rely on the observation of others to adapt their behaviors, revise their judgments, or make decisions [1–4]. In human populations, the access to social information has been greatly facilitated by the ongoing growth of communication technology. In fact, people are constantly exposed to a steady flow of opinions, advice and judgments of others about political ideas, new technologies, or commercial products [5]. When facing the opinions of peers on a given issue, people tend to filter and integrate the social information they receive and adjust their own beliefs accordingly [6,7]. At the scale of a group, repeated local influences among group members may give rise to complex patterns of opinion dynamics such as consensus formation, polarization, or fragmentation [8–11]. For example, it has been shown that people sharing similar extreme opinions, such as racial prejudices, tend to strengthen their judgment and confidence after interacting with one another [12]. Similar mechanisms of opinion dynamics can take place in a variety of social contexts, such as within a group of friends exchanging opinions about their willingness to get vaccinated against influenza [13,14]. At even larger scales, local influences among friends, family members, or coworkers — often combined with the global effects of mass media — constitute a major mechanism driving opinion formation during elections, shaping cultural markets [15], producing amplification or attenuation of risk perceptions [16,17], and shaping public opinion about social issues, such as atomic energy or climate change [18].

Given the remarkably large scope of social phenomena that are shaped by social influence and opinion dynamics, it is surprising that the behavioral mechanisms underlying these processes remain poorly understood. Important issues remain open: How do people adjust their judgment during social interactions? What are the underlying heuristics of opinion adaptation? And how do these *local* influences eventually generate *global* patterns of opinion change? Much of the existing modeling work about opinion dynamics has been addressed from a physics-based point of view, where the basic mechanisms of social influence are derived from analogies with physical systems, in particular with spin systems [19–23]. The wide variety of existing models assumes that individuals hold binary or continuous opinion values (usually lying between -1 and 1), which are updated over repeated interactions among neighboring agents. Different models assume different rules of opinion adaptation, such as imitation [24], averaging over people with similar opinions [25,26], following the majority [27], or more sophisticated equations [8,22]. Although informative as to the complex dynamics that can possibly emerge in a collective context, these simulation-based contributions share a common drawback: the absence of empirical verification of the models' assumptions [28]. Indeed, it is difficult to track and measure how opinions change under experimental conditions, as these changes depend on many social and psychological factors such as the personality of the individuals, their confidence level, their credibility, their social status, or their persuasive power [18]. In other disciplines such as psychology and cognitive science, laboratory experiments have been conducted to study how people integrate feedback from other individuals to revise their initial answers to factual questions [6,29,30]. However, the findings of local rules of opinion adaptation have not yet been used to study the *collective* dynamics of the system, and it remains unclear how social influence plays out in larger scale social contexts over time [31].

The present work draws upon experimental methods inspired by social psychology and theoretical concepts of complex systems typical of statistical physics. First, we conducted controlled experiments to describe the micro-level mechanisms of social influence, that is, how individuals revise their initial beliefs after being exposed to the opinion of another person. Then, we elaborated an individual-based model of social influence, which served to investigate the collective dynamics of the system. In a first experiment (see Materials & Methods), 52 participants were instructed to answer a series of 32 general knowledge questions and evaluate their confidence level on a scale ranging from 1 (*very unsure*) to 6 (*very sure*). This baseline experiment was used to characterize the initial configuration of the system before any social influence occurs. In a second experimental session, 59 participants answered 15 questions in the same way but were then exposed to the estimate and confidence level of another participant (henceforth referred to as "feedback") and asked to revise their initial answer. This procedure renders opinion changes traceable, and the effects of social influence *measureable* at the individual level. Moreover, changes in confidence were tracked as well, by asking participants to evaluate their confidence level before and after social influence. Despite empirical evidence suggesting that changes of opinion and confidence are intimately related [29], and theoretical work emphasizing the important role of inflexible, highly confident agents [32,33], this aspect of social influence remains poorly understood. Following the methods of existing experiments, we deliberately asked neutral, general knowledge questions, which allows capturing the mechanisms of opinion adaptation while controlling its emotional impact [6,30]. By exploring a simple model derived from our observations, we demonstrate that the collective dynamics of opinion formation in large groups of people are driven by two major "attractors of opinion": (*i*) the presence of a highly confident individual and (*ii*) the presence of clusters of low-confidence individuals sharing a similar opinion. In particular, we show that a critical amount of approximately 15% of experts is necessary to counteract the attractive effect of a large majority of lay individuals. As people are embedded in strongly connected social networks and permanently influence one another, these results constitute a first step toward a better understanding of the mechanisms of propagation, reinforcement, or polarization of ideas and attitudes in modern societies.

## *Results*

**Experimental results.** We first use the data from the first experiment to characterize the initial configuration of the system before any social influence occurs, that is, how opinions are initially distributed and how the accuracy and confidence of the answers are correlated with each other.

As shown in the example in Fig. 1A, the initial distribution of opinions has a lognormal shape, with a typical long tail indicating the significant presence of outliers. For each of 32 items we performed a Kolmogorov-Smirnov normality test of log($O_i$), where $O_i$ is the initial opinion of individual *i*. The test yielded p-values above .05 for 84% of the items, indicating that the null hypothesis cannot be rejected at the 5% significance level for these items. The remaining 16% still had reasonably high p-values (always >$10^{-3}$), suggesting that the initial opinions $O_i$ indeed follow a lognormal distribution.

We also analyzed the correlation between the confidence level of the participants and the accuracy of their answer (Fig. 1B). Interestingly, the confidence level is not such a reliable cue for accuracy [34]. First, we found no significant correlation between an individual *i*'s confidence level $C_i$ and the quality of his or her answer (a correlation test between $C_i$ and the error $Err(O_i) = \left|1 - \frac{O_i}{T}\right|$ where *T* is the true value yielded a coefficient of −.03). Nevertheless, a trend can be highlighted by grouping the data into classes of error ranges: very good answers ($Err(O_i) \leq 0.1$), good answers ($0.1 < Err(O_i) \leq 0.3$), bad answers ($0.3 < Err(O_i) \leq 1$) and very bad answers ($Err(O_i) > 1$). As it can be seen from Fig. 1B, only the maximum confidence level $C_i=6$ is a relevant indicator of the quality of the answer, leading to a good or very good estimate in 80% of the time. By contrast, lower confidence levels are less informative about accuracy. For instance, the second highest confidence value of $C_i$=5 has a 39% chance to correspond to a bad or very bad estimate. Similarly, a value of $C_i$=4 is more likely to accompany a bad or very bad estimate (53%) than a good or very good one (47%). The lowest confidence values $C_i$=1 and $C_i$=2 do not differ from each other. Taking the revised estimates of Experiment 2 into account, we observe that the reliability of high confidence judgments is undermined by social influence [29]. As shown in Fig. 2B, the distribution of errors for very confident individuals ($C_i$=5 or 6) becomes more noisy, widespread and clustered around certain values thus becoming less informative about accuracy after social influence.

To explore the wisdom of crowds, we compared the accuracy of various aggregating methods before and after social influence occurred (Fig. 2A). Our results agree with previous findings [29,35]. We find that the error distributions tend to become widespread, now covering a greater proportion of also high error values after social influence, regardless of the aggregating method.

Next, we focus on how people adjust their opinion after being informed about the opinion of another individual, which is the aim of Experiment 2. In agreement with previous studies [6,30], our results show that two variables have an important influence on how the individual *i* revises his or her opinion when exposed to the opinion and confidence of another participant *j*: the difference in confidence values $\Delta C_{ij} = C_i - C_j$ and the normalized distance between opinions: $\Delta O_{ij} = |O_j - O_i|/O_i$, where $O_j$ and $C_j$ represent the opinion and confidence level of participant *j*, respectively [6]. To provide a visual, quantitative overview of the effects of social influence, we draw an *influence map* that illustrates the interplay of these two variables in the process of opinion adaptation (Fig. 3). For the sake of simplicity, we distinguish three possible heuristics [30]:

1. *Keep initial opinion*, when individuals do not change their judgment after receiving a feedback, that is: $R_i = O_i$, where $R_i$ is the revised opinion of participant *i*.

2. *Make a compromise*, when the revised opinion falls in between the initial opinion $O_i$ and the feedback $O_j$: min($O_i$, $O_j$)< $R_i$ <max($O_i$, $O_j$).

3. *Adopt other opinion*, when an individual *i* adopts the partner's opinion: $R_i = O_j$.

The influence map shows the heuristic that is used by the majority of people as $\Delta C_{ij}$ and $\Delta O_{ij}$ change (Fig. 3A). Most of the data points (86% of 885) are found for $-3 \leq \Delta C_{ij} \leq 3$ and $\Delta O_{ij} \leq 1.2$, which cover a large part of the influence map and seem to be reasonable ranges being also encountered in real life situations. At the edge of the map, however, the results are more uncertain due to the scarcity of available data points.

Figure 3A shows that the first and more conservative strategy tends to dominate the two others. In particular, the majority of people systematically keep their opinion when the value of $\Delta C_{ij}$ is positive, that is, when their own confidence exceeds their partner's [30]. However, when their confidence level is equal or lower than their partner's, individuals tend to adapt their opinion accordingly. Importantly, one can distinguish three zones in the influence map, according to the distance between estimates $\Delta O_{ij}$ (Fig. 3B). First, when both individuals have a similar opinion ($\Delta O_{ij} < 0.3$), individuals tend to keep their initial judgment, irrespective of their partner's confidence. Moreover, they also have a strong tendency to increase their confidence level (see Fig. 4A indicating the changes in confidence). Therefore, we interpret this area of agreement as being a *confirmation zone*, where feedback tends to simultaneously reinforce initial opinions and increase an individual's confidence.

It turns out that feedback has the strongest influence at intermediate levels of disagreement, when $0.3 < \Delta O_{ij} < 1.1$. In this zone, the "*compromise*" heuristic is selected by most people when $-3 \leq \Delta C_{ij} \leq 0$, and the "*adoption*" heuristic appears for lower values of $\Delta C_{ij}$. We call this the *influence zone*, where social influence is strongest. Here, the other's opinion differs sufficiently from the initial opinion to trigger a revision but is still not far enough away to be completely ignored. In particular, the confidence level of the participants tends to remain the same after the interaction (Fig. 4B).

Finally, when the distance between opinions is very large (i.e., $\Delta O_{ij} > 1$), the strength of social influence diminishes progressively [6]. In this zone, people seem to pay little attention to the judgment of another, presumably assuming that it may be an erroneous answer. Nevertheless, the other's opinion is not entirely ignored, as the majority of people still choose the "*compromise*" heuristic when the partner is markedly more confident (i.e. $\Delta C_{ij} \leq 2$). Moreover, people who are initially very confident (i.e. $C_i \geq 5$) presumably begin to doubt the accuracy of their judgment and exhibit a high likelihood (of almost 70%) of reducing their confidence level. Even more remote opinions are likely to be ignored entirely, but as this situation rarely occurs our data does not warrant a reliable conclusion here.

**The model.** Taking these empirical regularities into account, we now elaborate an individual-based model of opinion adaptation and explore the collective dynamics of opinion change when many people influence each other repeatedly. To this end, we first describe the above influence map by means of a simplified diagram showing the heuristics that are used by most individuals according to $\Delta O_{ij}$ and $\Delta C_{ij}$ (Fig. 3B). Alternatively, the same diagram can be characterized as a decision tree (Fig. 3C). The model is defined as follows:

First, an individual notes the distance $\Delta O_{ij}$ between his or her own and a partner's opinion and classifies it as near, far, or at an intermediate distance. For this, we used two threshold values of $\tau_1 = 0.3$ and $\tau_2 = 1.1$, assuming that the feedback is near when $\Delta O_{ij} < \tau_1$, far when $\Delta O_{ij} > \tau_2$, and at an intermediate distance otherwise. The numerical values of $\tau_1$ and $\tau_2$ were determined empirically from the influence map. Second, the individual considers the difference in confidence values $\Delta C_{ij}$ to choose among the three heuristics. Again, we define two threshold values $\alpha_1$ and $\alpha_2$ and assume that the individual decides to "*keep own opinion*" if $\Delta C_{ij} \geq \alpha_1$, to "*adopt other opinion*" if $\Delta C_{ij} \leq \alpha_2$, and to "*make a compromise*" otherwise. The three strategies can be formally defined as $R_i = O_i + \omega(O_j - O_i)$, where the parameter $\omega$ delineates the strength of social influence. Therefore, we have $\omega = 0$ when the individual decides to "*keep own opinion*", and $\omega = 1$ when the individual decides to "*adopt*". When the individual chooses the "*compromise*" strategy, that is when $0 > \omega > 1$, the average weight value $\bar{\omega}$ as measured from our data equals to $\bar{\omega} = 0.4$ (*SD*=0.24), indicating that people did not move exactly between their initial estimate and the feedback (which would correspond to a weight value of 0.5), but exhibited a bias toward their own initial opinion [30]. Over all our data points, 53% correspond to the first strategy ($\omega = 0$), 43% to the second ($0 > \omega > 1$), and 4% to the third ($\omega = 1$).

The values of $\alpha_1$ and $\alpha_2$ depend on the distance zone defined before:

- When $\Delta O_{ij}$ is small, the other's opinion constitutes a *confirmation* of the initial opinion. According to our observations, $\alpha_1$=-5 and $\alpha_2$=-6. Additionally, the confidence level $C_i$ is increased by one point if $\Delta C_{ij} \leq -4$. As indicated by Fig. 4A, $C_i$ is also increased by one point with a probability $p$=0.5 when $-4 \leq \Delta C_{ij} \leq 0$, and remains the same otherwise.

- When $\Delta O_{ij}$ is intermediate, the feedback has a significant influence on the subject's opinion. In this case, we set $\alpha_1$=0 and $\alpha_2$=-3. The data shows that the confidence level is changed only if $\Delta C_{ij} \leq -3$ (Fig. 4B). In this case, $C_i$ increases with probability $p$=0.5, and remains the same otherwise.

- When $\Delta O_{ij}$ is large, the thresholds are set to $\alpha_1$=-2 and $\alpha_2$=-6. This time, the confidence level decreases by one point when $\Delta C_{ij} \geq 4$, and remains the same otherwise.

Here, all the parameter values were directly extracted from the observations (Fig.3B and Fig.4).

**Collective dynamics.** Having characterized the effects of social influence at the individual level, we now scale up to the collective level and study how *repeated* influences among *many* people play out at the population scale. Because the macroscopic features of the system are only visible when a large number of people interact many times, it would be extremely difficult to investigate this under laboratory conditions. Therefore, we conducted a series of numerical simulations of the above model to investigate the collective dynamics of the system.

The initial conditions of our simulations correspond to the exact starting configurations observed in our experiments (i.e., the precise opinion and confidence values of all 52 participants observed in the first experiment) [36]. In each simulation round, the 52 individuals are randomly grouped into pairs, and both individuals in a pair update their opinions according to the opinion of the other person, as predicted by our model. Thus, each individual is both a source and the target of social influence. We performed $N$=300 rounds of simulated interactions, where $N$ has been chosen large enough to ensure that the system has reached a stationary state. Here, we make the assumption that the decision tree that has been extracted from our experiment remains the same over repeated interactions. This assumption is reasonable to the extent that the outcome of the decision tree (i.e. the strategy that is chosen) depends on the confidence level of the individual, which is expected to change as people receive new feedback. In such a way, the strategies that will be selected by individuals are connected to the individual history of past interactions.

Fig. 5 shows the dynamics observed for three representative examples of simulations. Although a certain level of opinion fragmentation still remains, a majority of individuals converge toward a similar opinion. As shown by the arrow maps in Fig.5, the first rounds of the simulation exhibit important movements of opinions among low-confidence individuals (as indicated by the large horizontal arrows for confidence lower than 3), without increase of confidence (as shown in Fig. S2). After a certain number of rounds, however, a tipping point occurs at which a critical proportion of people meet up in the same region of the opinion space. This creates a subsequent increase of confidence in this zone, which in turn becomes even more attractive to others. This results in a positive reinforcement loop, leading to a stationary state in which the majority of people end up sharing a similar opinion. This amplification process is also marked by a sharp transition of the system's global confidence level (Fig. S2), which is a typical signature of phase transitions in complex systems [2].

An intriguing finding of our simulations is that the collective opinion does not converge toward the average value of initial opinions (a correlation test yields a nonsignificant effect with a coefficient $c=-.05$). The correlation between the convergence point and the median value of the initial opinions is significant ($p=.03$) but the relatively moderate correlation coefficient $c=0.46$ suggests that this relation remains weak. Likewise, the system does not systematically converge toward or away from the true value (nonsignificant effect with a coefficient $c=.11$). Instead, the simulations exhibit complex collective dynamics in which the combined effect of various elements can drive the group in one direction or another. In agreement with previous works [15], the collective outcome appears to be poorly predictable and strongly dependent on the initial conditions [36]. Nevertheless, we identified two major *attractors of opinions* that exert an important social influence over the group:

1. The first attractor is the presence of a critical mass of uncertain individuals who happen to share a similar opinion. In fact, when such a cluster of individuals is initially present in the system—either by chance or because individuals share a common bias—the rest of the crowd tends to converge toward it, as illustrated by Fig. 5-Example2. This *majority effect* is typical of conformity experiments that have been conducted in the past [37], where a large number of people sharing the same opinion have a strong social influence on others.

2. The second attractor is the presence of one or a few highly confident individuals, as illustrated by Fig. 5-Example3. The origin of this *expert effect* is twofold: First, very confident individuals exert strong persuasive power, as shown by the influence map.

Second, unconfident people tend to increase their own confidence after interacting with a very confident person, creating a basin of attraction around that person's opinion [38,39].

Our simulations show that the *majority effect* and the *expert effect* are not systematically beneficial to the group, as both attractors could possibly drive the group away from the truth (Fig. 5-Example 2). What happens in the case of conflicting interests, when the expert and the majority effects apply simultaneously and disagree with each other (Fig. 5-Example 3)? To investigate this issue, we conducted another series of simulations in which a cluster of low-confidence individuals sharing the same opinion $O_{maj}$, is facing a minority of high-confidence experts holding another opinion $O_{exp}$. As shown by Fig. 6A, the majority effect overcomes the expert effect when the proportion of experts $p_{Exp}$ is lower than a certain threshold value located around 10%. However, as $p_{Exp}$ increases from 10%, to 20% a transition occurs and the convergence point shifts from the majority to the experts' opinion. Remarkably, this transition point remains stable even when a proportion $p_{Neut}$ of neutral individuals (defined as people with random opinions and a low confidence level) are present in the system (Fig. 6B). As $p_{Neut}$ increases above 70%, however, noise gradually starts to dominate, leading the expert and the majority effects to vanish. The tipping point occurring at a proportion of around 15% of experts appears to be a robust prediction, not only because it resists to a large amount of system noise (Fig. 6B), but also because a previous theoretical study using a completely different approach also reached a similar conclusion [40].

## *Discussion*

In this work, we have provided experimental measurements and quantitative descriptions of the effects of social influence—a key element in the formation of public opinions. Our approach consisted of three steps: using controlled experiments to measure the effects of social influence at the scale of the individual, deriving a simple process model of opinion adaptation, and scaling up from individual behavior to collective dynamics by means of computer simulations.

The first result of our experiment is that participants exhibited a significant bias toward their own initial opinion rather than equally weighting all social information they were exposed to [6,30]. This bias is visible from the influence map shown in Fig. 3, where the blue color corresponding to "*keep initial opinion*" is dominant and the red one corresponding to "*adopt the other opinion*" is rare. As shown in Fig. 3B, the same trend has been transferred to the model. Moreover, even when the "*compromise*" strategy is chosen, individuals still give a stronger weight $\omega = 0.4$ to their own initial opinion, which has also been implemented in the model. Therefore, contradictory feedback is typically underestimated—if not completely ignored—but opinions corroborating one's initial opinion trigger an increase in confidence. This observation is consistent with the so-called *confirmation bias* in psychology, namely, the tendency of people to pay more attention to information confirming their initial beliefs than information they disagree with [41,42]. This result is also in line with early experiments showing that opinions tend to get reinforced by group discussions that involve people who initially share a similar judgment [12]. Likewise, the fact that individuals holding completely different beliefs exert very little influence on each other is consistent with the idea of "bounded confidence"—a modeling concept suggesting that social influence is negligible when opinions are initially too distant [20,26]. The presence of these elements confirms that our experimental design has indeed captured the fundamental mechanisms of social influence, and that factual questions can be used, to some extent, to study the fundamental features of opinion dynamics [29]. In the future, an important challenge will be to evaluate how the influence map is shaped when emotions and subjective beliefs are more relevant (e.g. by using items about political opinions or beliefs that elicit strong convictions or emotions). Besides, another important follow-up study that should be conducted in the near future is the verification of our assumption that the decision tree observed at the first round of interaction remains identical over repeated interactions.

Scaling up from individual to collective behavior was achieved by means of computer simulations in line with existing approaches in the field of self-organization and complex systems [2,9,19]. Our simulations allowed us to unravel the precise mechanisms of opinion dynamics in large groups of people, which would have been practically impossible to characterize under laboratory conditions. In particular, an important ingredient underlying the collective dynamics but lacking in previous modeling approaches is the specific interplay between opinion changes and confidence changes. First, confidence serves as a sort of system memory. In fact, over simulation rounds, individuals are less easily influenced by others because their confidence level gradually increases as they receive new feedback. Therefore, simulated individuals do not constantly change their opinion but progressively converge toward a stable value in a realistic manner. Second, the increase of confidence supports the emergence of basins of attraction during collective opinion dynamics by boosting the attractive power of individuals sharing a similar opinion [29]. This process often turns out to be detrimental to the group, because the local amount of confidence may grow *artificially* in a given region of the opinion space, which provides false cues to others and triggers a snowball effect that may drive the group in an erroneous direction. Interestingly, judgments of high confidence are good indicators of accuracy before social influence occurs, but no longer after people have been exposed to the opinion of others. It is remarkable that even a mild influence has a significant impact on the reliability of high confidence cues, as shown in Fig. 2B. The main problem induced by social influence is that people tend to become more confident after noticing that other people have similar opinions. Therefore, high confidence is an indicator of accuracy when judgments are independent but becomes an indicator of *consensus* when social influence takes place [43,44].

Our simulation results also identified two elements that can cause such amplification loops: the expert effect—induced by the presence of a highly confident individual, and the majority effect—induced by a critical mass of low-confidence individuals sharing similar opinions. Moreover, the presence of a significant number of neutral individuals holding a random opinion and a low confidence level around these two attractive forces tends to increase the unpredictability of the final outcome [15]. Therefore, neutral individuals make the crowd less vulnerable to the influence of opinion attractors, and thus less predictable. By contrast, recent studies on animal groups have shown that the presence of uninformed individuals in fish schools acts in favor of the numerical majority, at the expense of very opinionated individuals [1].

Our simulations constitute a valuable tool that allows (*i*) unravelling the underlying mechanisms of the system, (*ii*) forecasting future trends of opinion change, and (*iii*) driving further experimental research and data collection. Nevertheless, it is important to note that the outcome of our simulations requires empirical validation in the future. This could be addressed, for instance, by means of empirical observations over the Web, where one would measure people's opinion about a social issue over blogs and discussion forums and evaluate how the collective opinion changes over time [45,46]. Alternatively, an online experimental approach such as the one elaborated by Salganik et al. seems well suited to the study of opinion dynamics under controlled conditions [15].

By quantifying the balance of power between the expert effect, the majority effect, and neutral individuals, our research can inform applications regarding the management of situations in which a small opinionated minority challenges a large population of uninformed individuals. For example, the model could help doctors convince a population of laypeople to adopt certain disease prevention methods or reversely prevent extremist groups from taking control of a large group of people.

## *Materials and Methods*

**Ethics Statement.** The present study has been approved by the Ethics Committee of the Max Planck Institute for Human Development. All participants gave written and informed consent to the experimental procedure.

**Experimental design.** The experimental part of the study consisted of two distinct experiments: one without social influence (Experiment 1) and one with (Experiment 2). In both experiments, participants entered the laboratory individually and were instructed to answer a series of factual questions displayed on a computer screen. All participants were naïve to the purpose of our experiments and received a flat fee of €8. In Experiment 1, a total of 52 participants ($M_{age}$=27 years, SD=9, 50% females) responded to 32 general knowledge questions, which covered the areas sports, nature, geography and society/economy (8 per area; for a complete list of items see Table S1). The correct answers to the questions ranged from 100 to 999, which, however, was not known to the participants. Participants were instructed to respond as accurately as possible and to indicate their confidence on a 6-point Likert scale (1 *very unsure* to 6 *very sure*) after having given their spontaneous estimate. Questions were displayed one after the other on the computer screen, and a new question was given only after participants answered the current one. Participants were only informed about the correct answers to the questions after the end of the experiment and therefore could not figure out that the true values always lied in the interval [100 999]. The order of the questions was randomized for each participant. A correlation test of the accuracy of answers and the order of the questions yielded non-significant p-values for 90% of participants with a probability p>0.05, confirming the absence of any learning process over experimental rounds. After the end of the experiments, participants were paid, thanked and released. In Experiment 1, participants were not exposed to the social influence of others. The 1664 data points (corresponding to 52 participants × 32 questions) were used to characterize the features of the initial environment, such as the distribution of answers and the analyses of the confidence levels shown in Fig. 1, and as a pool of social influence for the second experiment. The same dataset was used to define the initial condition of the simulations presented in Fig. 5.

In Experiment 2, 59 participants ($M_{age}$=33 years, SD=11, 56% females) responded to 15 of the 32 general knowledge questions used in Experiment 1 and indicated their confidence level. Experiment 2 was conducted under the same conditions as in Experiment 1 except that participants were informed that they would receive a feedback from another participant. After each question, the estimate and confidence level of another randomly selected participant from Experiment 1 were displayed on the computer screen, and participants were then asked for a revised estimate and corresponding confidence level. This second dataset made of 59×15=885 binary interactions was used to study the effects of social influence, from which we derived the results shown in Figs. 2, 3 and 4. The full list of questions is available in Table S1.

## Acknowledgments

We are grateful to Gudrun Rauwolf and Tor Nielsen for their inspiring feedbacks and participation in the work, and Isaac and Jeanne Gouëllo for fruitful discussions. We thank Jack Soll for sharing with us the experimental data from ref. [30], and two anonymous reviewers for their constructive comments. The authors thank Anita Todd for language editing.

**Figures**

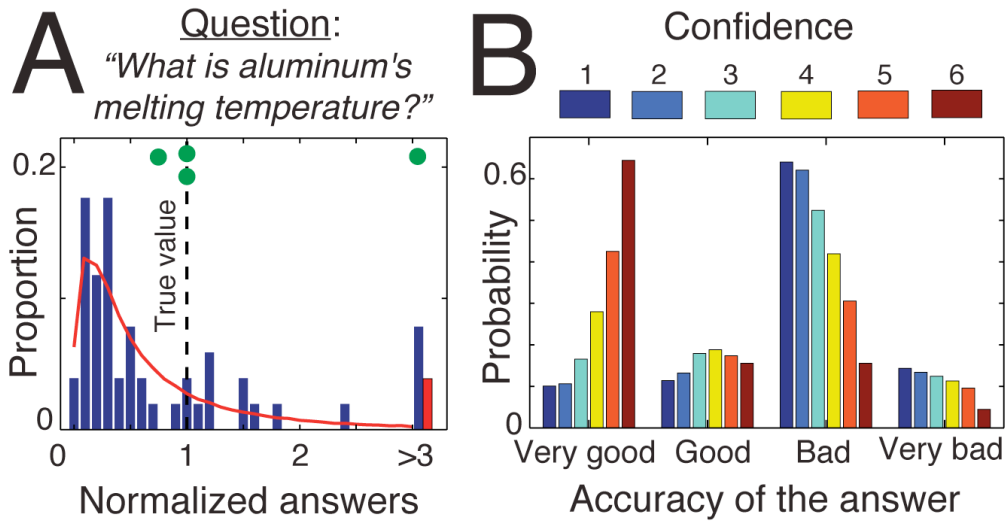

**Figure 1:** The initial configuration of the system in the absence of social influence. (A) Initial distribution of opinions for one representative example question (see Fig. S1 for an overview of all 32 items). The normalized answer corresponds to the estimate of the participants divided by the true value (i.e., 660°C for this question). The red curve shows the best fit of a lognormal distribution. The green dots at the top indicate the location of estimates associated with high confidence levels ($C_i \geq 5$). One of them constitutes an outlier. (B) Accuracy of participants' answers as a function of their confidence level, as determined from the complete dataset (32 items).

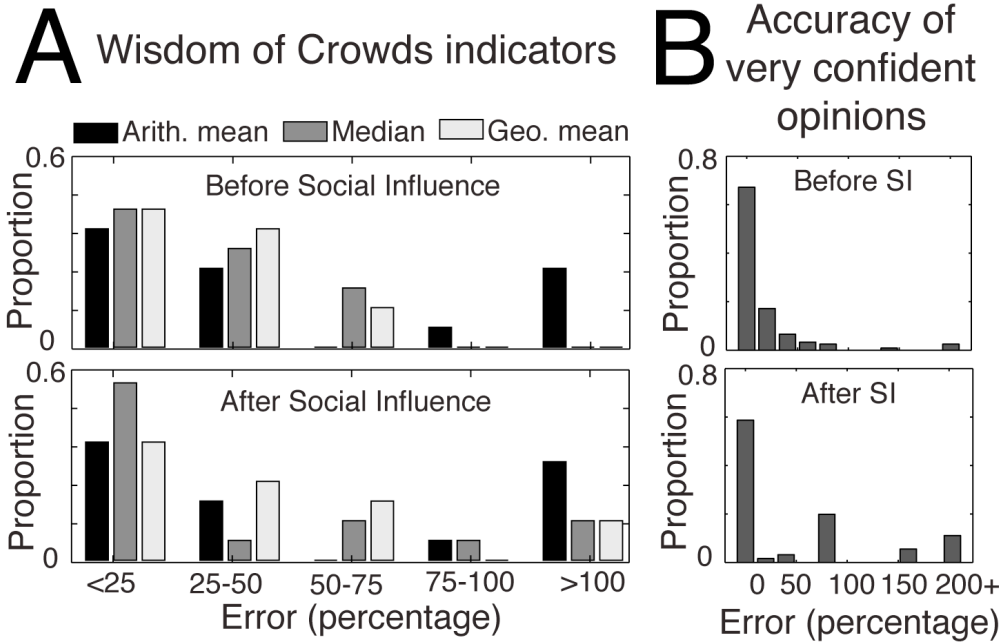

**Figure 2:** Effects of social influence on the wisdom of crowds (A), and the relevance of the confidence cue (B). The error is the deviation from the true value as a percentage. (A) Before any social influence occurs, the arithmetic (Arith.) mean is sensitive to single extreme opinions and does not appear as a relevant aggregating method. The median and geometric (Geo.) mean are more robust to outliers. When social influence occurs, however, the distributions are skewed to the right and the three indicators are more likely to generate high error values. (B) In the absence of social influence (SI), a clear and continuous trend is visible, where individuals with high confidence ($C_i \geq 5$) constitute a good indicator of the quality of the answer. When social influence is injected in the system, however, the distribution becomes noisier and less predictable. Overall, social influence generates unpredictability in the observed trends.

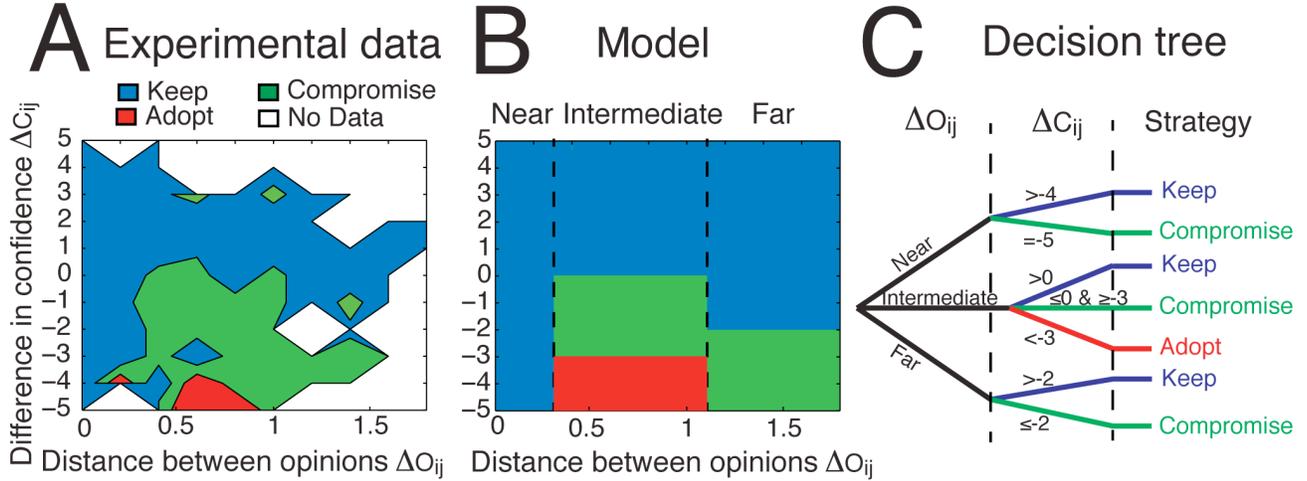

**Figure 3**: (A) The influence map extracted from our experimental data and (B) a simplified representation of it as implemented in the model. The color coding indicates the heuristic that is used by a majority of people, as a function of the difference in confidence $\Delta C_{ij} = C_i - C_j$ and the distance between the normalized opinions $\Delta O_{ij} = |O_j - O_i|/O_i$. Positive values of $\Delta C_{ij}$ indicate that the focus subject is more confident than the influencing individual (called feedback), whereas negative values indicate that the focus subject is less confident. White zones in (A) indicate the absence of sufficient data. Although the majority of people prefer to keep their initial opinion when they are more confident than their partner (i.e. the blue strategy dominates for $\Delta C_{ij} > 0$), a zone of strong influence is found at an intermediate distance with $\Delta C_{ij} < 0$. (C) The decision tree describing the decision process with three different outcome strategies. The individual first looks at the distance between opinions $\Delta O_{ij}$, then looks at the difference of confidence $\Delta C_{ij}$, and finally chooses a strategy accordingly.

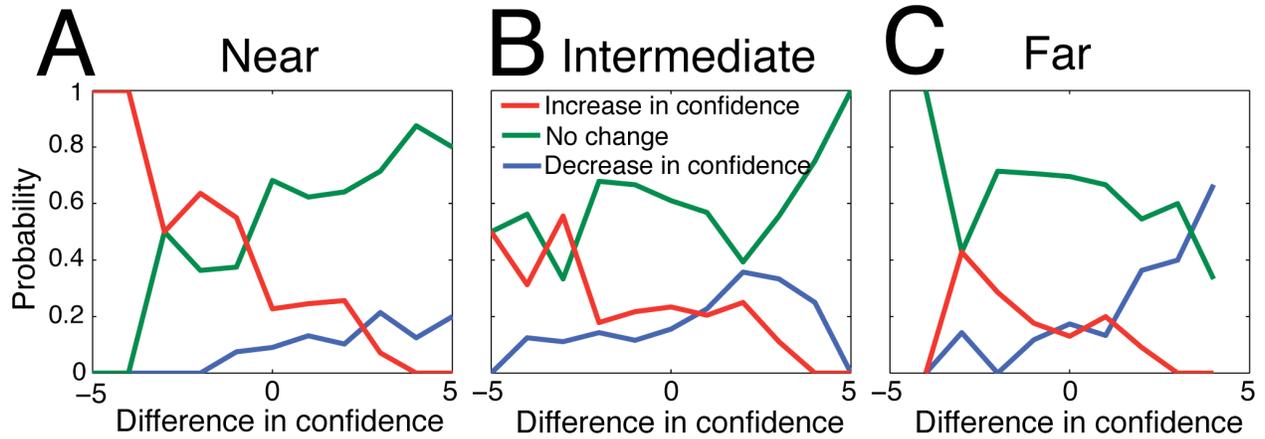

**Figure 4**: The probability of increasing (red), decreasing (blue), or maintaining (green) the confidence level after social influence. Changes in confidence are indicated according to the opinion distance classes as defined in the influence map (Fig. 3): (A) *near* when $\Delta O_{ij} \leq 0.3$, (B) *intermediate* when $0.3 < \Delta O_{ij} \leq 1.1$, and (C) *far* when $\Delta O_{ij} > 1.1$. A tendency to increase confidence is visible in the *near* and *intermediate* zones when participants interact with a more confident subject. Confidence can also decrease in the *far* zone, when $\Delta C_{ij} \geq 4$.

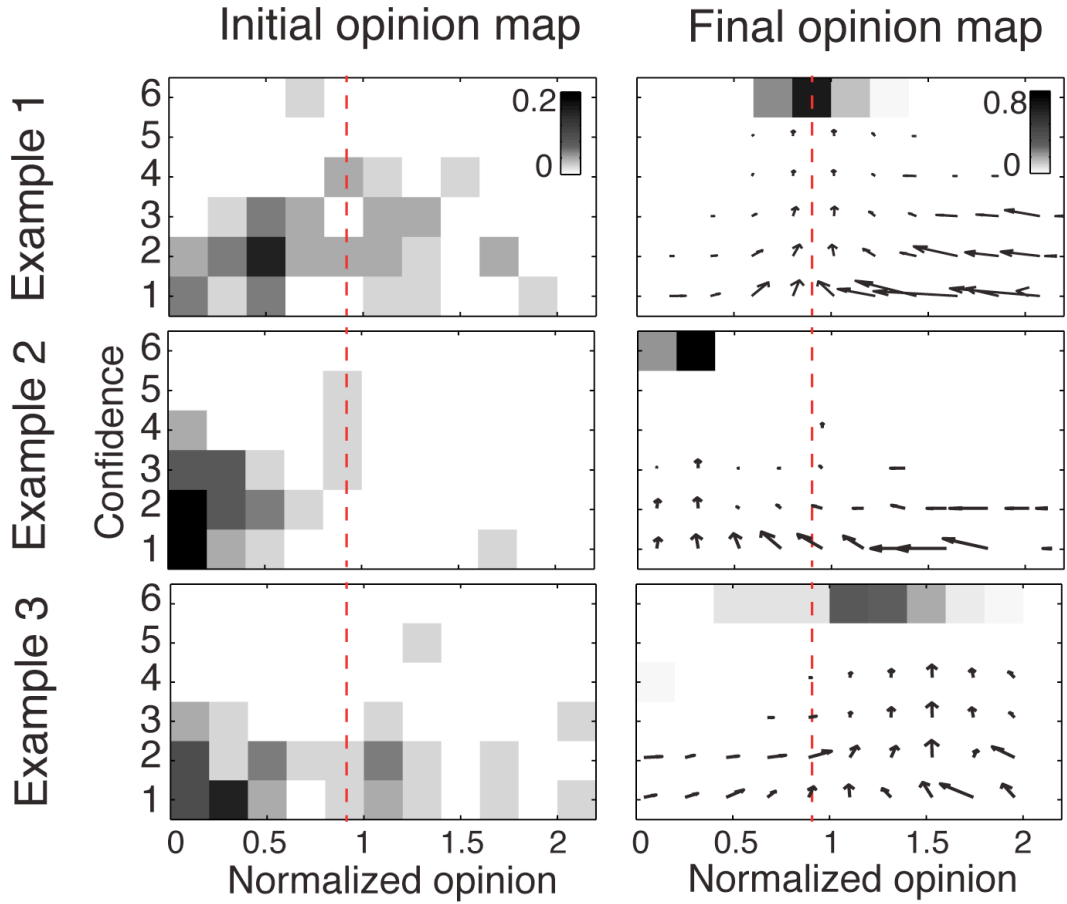

**Figure 5:** Three representative examples of the collective dynamics observed in the computer simulations. For each example, the initial opinion map is shown on the left-hand side (experimental data), and the final opinion map after $N=300$ rounds of simulations on the right-hand side. The opinion maps represent the proportion of individuals with a given opinion (*x*-axis) and a given confidence level (*y*-axis). As in Fig. 1, the normalized opinion is the actual opinion divided by the true value. The correct answer is represented by the red dashed lines (corresponding to a value of 1). Outliers with normalized opinion greater than 2 are not shown. The arrow maps represent the average movements over both opinion and confidence dimensions during simulations. Examples 1, 2, and 3 correspond to the questions "*What is the length of the river Oder in kilometers?*", "*How many inhabitants has the East Frisian island Wangerooge?*", and "*How many gold medals were awarded during the Olympics in China in 2008?*", respectively. The final convergence point may be determined by a dense cluster of low confidence individuals, as illustrated by Example 2 (*majority effect*), or by a few very confident individuals as in Example 3 (*expert effect*).

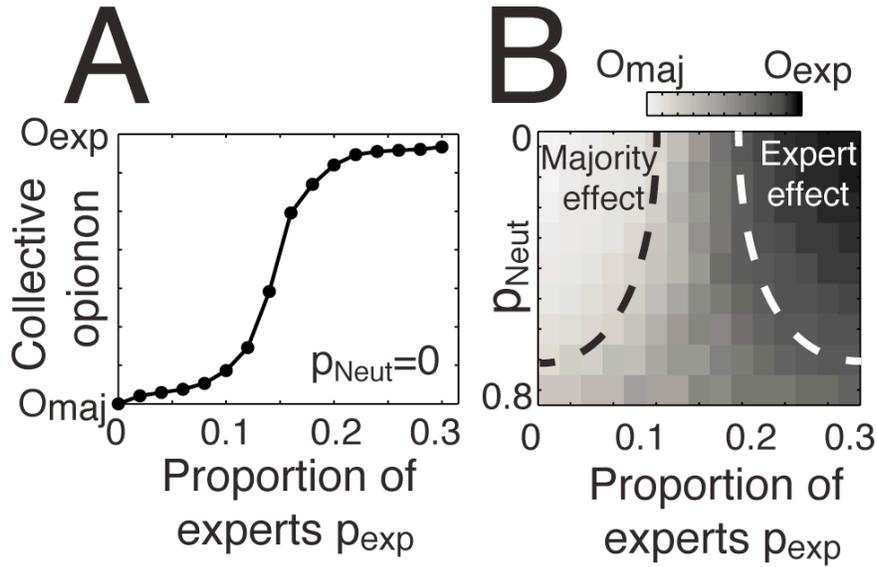

**Figure 6**: Which attractor dominates when the majority effect and the expert effect apply simultaneously? (A) The evolution of collective opinion when varying the relative proportion of experts $p_{Exp}$, holding an opinion $O_{exp}$ and a high confidence level $C_{exp}=6$, and the proportion of people in the majority group $p_{maj}$ holding an opinion $O_{maj}$ and a low confidence level randomly chosen in the interval $C_{maj}=[1\ 3]$. Here, the number of neutral individuals is fixed to $p_{Neut}=0$. (B) Phase diagram showing the parameter space where the majority or the expert effects applies, when increasing the proportion of neutral individuals $p_{Neut}$ holding a random opinion and a low confidence level randomly chosen in the interval $C_{uni}=[1\ 3]$. The schematic regions delimited by black or white dashed lines show the zones where the collective opinion converges toward the majority or the expert opinion, respectively. In the transition zone, the collective opinion converges somewhere between $O_{exp}$ and $O_{maj}$. In some rare cases, the crowd splits into two groups or more.

## Supporting Information Legends

**Figure S1**: The distribution of answers for all 32 questions used in the first experiment (Experiment1, see Materials & Methods). The numbers on the upper right corner correspond to the question *id*, as indicated in the list of questions provided in the table S1. Question *id*=27 has been used for illustrative purpose in the main text (Fig. 1A). The normalized answer is the estimate of the participants divided by the true value. The black dashed lines indicate the correct answer (normalized value = 1). The red and green dashed lines indicate the mean and the median values of the distribution, respectively. The mean values lying farther than 3 are not indicated.

**Figure S2**: Three representative examples showing the evolution of participants' confidence over simulation rounds. Examples 1, 2 and 3 correspond to those shown in Fig. 4 in the main text. The average global confidence is computed by taking the mean value of confidence for all 52 participants. After a few rounds of simulation, a sharp transition occurs toward high confidence levels, attesting for the opinion amplification process.

**Table S1**: Full list of questions used in the study.